\documentclass[11pt]{article}
\usepackage[utf8]{inputenc} 
\usepackage{amssymb, amsthm, amsmath} 
\usepackage{geometry} 
\usepackage{graphicx} 
\usepackage{booktabs} 
\usepackage{array} 
\usepackage{authblk}
\usepackage{paralist} 
\usepackage{verbatim} 
\usepackage{subfig} 
\usepackage{fancyhdr}
\usepackage{indentfirst}
\usepackage{sectsty}
\allsectionsfont{\sffamily\mdseries\upshape}
\usepackage[nottoc,notlof,notlot]{tocbibind} 
\usepackage[titles,subfigure]{tocloft}

\newcommand{\F}{\mathbb F}
\newcommand{\Tr}{\mathbf{Tr}_{\F_{p^{ms}} \setminus \F_{p^s}}}
\newcommand{\Trsix}{\mathbf{Tr}_{\F_{2^{8}} \setminus \F_{2^4}}}

\newcommand{\N}{\mathbf{N}_{\F_{q^m} \setminus \F_q}}
\newcommand{\Nsix}{\mathbf{N}_{\F_{2^8} \setminus \F_{2^4}}}

\newtheorem{definition}{Definition}
\newtheorem{proposition}{Proposition}

\makeatletter
\def\thanks#1{\protected@xdef\@thanks{\@thanks
        \protect\footnotetext{#1}}}
\makeatother

\title{Introducing Three Best Known Binary Goppa Codes}

\author[1]{Jan L. Carrasquillo--L\'opez}
\author[2]{Axel O. G\'omez--Flores}
\author[3]{Christopher Soto}
\author[4]{Fernando L. Pi\~nero Gonz\'alez}
\affil[1]{Department of Mathematics, University of Puerto Rico at Cayey}
\affil[2]{Department of Mathematics, University of Puerto Rico at R\'io Piedras}
\affil[3]{Department of Mathematics, Queens College of the City University of New York}
\affil[4]{Department of Mathematics, University of Puerto Rico in Ponce}
\affil[ ]{\textit {jan.carrasquillo1@upr.edu,axel.gomez@upr.edu,christopher.soto32@qmail.cuny.edu, fernando.pinero1@upr.edu  }}

\thanks{Supported by NSF-DMS REU-1852171.}

\begin{document}

\maketitle

\begin{abstract}
\noindent The current best known $[239, 21], \, [240, 21], \, \text{and} \, [241, 21]$ binary linear codes have minimum distance 98, 98, and 99 respectively. In this article, we introduce three binary Goppa codes with Goppa polynomials  $(x^{17} + 1)^6, (x^{16} + x)^6,\text{ and } (x^{15} + 1)^6$. The Goppa codes are $[239, 21, 103], \, [240, 21, 104], \, \text{and} \, [241, 21, 104]$ binary linear codes respectively. These codes have greater minimum distance than the current best known codes with the respective length and dimension. In addition, with the techniques of puncturing, shortening, and extending, we find more derived codes with a better minimum distance than the current best known codes with the respective length and dimension.
\end{abstract}

\section{Introduction}
Binary Goppa codes have been one of the most widely studied linear codes since V. D. Goppa introduced them in 1970 \cite{Goppa-70}. Due to their algebraic structure and good decoding capabilities, they are important in many applications, for example, telecommunications and post-quantum cryptography systems such as the McEliece cryptosystem. The McEliece cryptosystem was introduced in 1978 \cite{M-78} and incorporates the use of binary Goppa codes in its original algorithm; the cryptosystem has resisted most cryptographic attacks.

Some of the best known linear codes are constructed from binary Goppa codes. In this article, we propose three new binary Goppa whose parameters are better than what is currently know. These binary Goppa codes have Goppa polynomials: 
\[g(x) = (x^{17} + 1)^6 , (x^{16} + x)^6, \text{ and } (x^{15} + 1)^6.\]

\section{Background}

\begin{definition}\cite{TEXTBOOK}
Given a linear code $C$ with length $n$, let $A_w$ denote the number of codewords whose weight equals w. Then, the vector $[A_0, A_1, ..., A_n]$ is called the weight enumerator of $C$. The weight enumerator polynomial of $C$ is defined by
$$
 W(C ; x , y) = \sum_{w = 0}^n A_w x^w y^{n - w}.
$$
\end{definition}




\noindent The lowest positive weight $w$ such that $A_w \neq 0$ is the minimum distance of the code. \\

\begin{definition} \cite{Goppa-70}
Let $p$ be a prime and let $q = p^m$. Let $L = \{ \alpha_1, \alpha_2, \ldots, \alpha_n\}$ be a subset of $\F_q$. Let $g(x)$ be a polynomial of degree $t$ such that $g(\alpha_i) \neq 0 $, for $\alpha_i \in L$. The \emph{$p$--ary Goppa code} is defined as
$$
 C(L, g) := \left\{ (c_1, c_2, \ldots, c_n) \in \F_{p}^n  \ | \ \sum_{i = 1}^n \frac{c_i}{x-\alpha_i} \equiv 0 \pmod{g(x)} \right\}.
$$ 
\end{definition}

The exact computation of the dimension and the minimum distance of a given Goppa code remain an open problem. The following bound on the parameters of binary Goppa codes is well known:

\begin{proposition}\cite{Goppa-70}
Let $p$ be a prime power and let $q = p^m$. Let $L = \{ \alpha_1, \alpha_2, \ldots, \alpha_n\}$ be a subset of $\F_q$. Let $g(x)$ be a polynomial of degree $t$ such that $g(\alpha_i) \neq 0 $, for $\alpha_i \in L$.  Then the dimension  of $C(L, g)$ is at least $n - mt$ and the minimum distance of $C(L, g)$ is at least $t + 1$.
\end{proposition}

The following identity involving Goppa codes turns out to be crucial to improve the minimum distance bounds.



\begin{proposition}\label{prop:GoppaBound2}\cite{SKHN-76}
Let $g$ be an irreducible polynomial. Then the $p$--ary Goppa code satisfies
$$
 C(L, g^{ap-1}) =  C(L,g^{ap}).
$$
\end{proposition}

Goppa codes are subfield subcodes of Generalized Reed–Solomon codes. They form a subfamily of alternant codes that provides lower bounds for their dimension and minimum distance. Sometimes, these bounds may be improved, depending on the choice of the Goppa polynomial. For example,  P. Verón improved the dimension bound for binary Goppa codes whose Goppa polynomial is given by $g(x) = a(x)\Tr(b(x))$, where $a(x)$ and $b(x)$ are polynomials. Recently, other improvements have been made to the dimension and minimum distance bounds of $q$-ary Norm Goppa codes ($q = p^s$), i.e. Goppa codes over $\mathbb{F}_q$ where the Goppa polynomial is $\N(b(x)) = b(x)^{1+q+q^2 +\cdots q^{m-1}}$ \cite{COT-14}. Our Goppa polynomials are related to these classes codes because  $x^{16}+x = \Trsix(x)$,  $x^{15}+1  = \dfrac{\Trsix(x)}{x}$, and $x^{17}+1 = \Nsix(x) +1$.

\section{Results}

We have used the Coding Theory library of the SageMath programming language \cite{SAGE} to determine the parameters of our codes. In particular, we used SAGE's own Goppa Codes constructor and its method to compute the weight distribution of each Goppa code.



\subsection{Construction of $[239,21,103]$ code:}
We computed that the binary Goppa code $C\left(L,(x^{17}+1)^6\right)$ is a $[239, 21, 103]$ code. This linear code has a higher minimum distance than the current best known $[239, 21, 98]$ binary code. Its weight enumerator polynomial is given by:
\begin{align*}
    x^{239} &+ 62244 x^{136} y^{103} + 81396 x^{135} y^{104} + 190519 x^{128} y^{111} + 217736 x^{127} y^{112} \\
    &+ 496680 x^{120} y^{119} + 496680 x^{119} y^{120} + 217736 x^{112} y^{127} + 190519 x^{111} y^{128} \\ 
    &+ 81396 x^{104} y^{135} + 62244 x^{103} y^{136} + y^{239}.
\end{align*}

The $[239,21,103]$ binary Goppa code $C(L, (x^{17}+1)^6)$ is related to the $[239,123,35]$ binary Goppa code $C(L, x^{17}+1)$ (a  best known binary code) and the $[55,16,19]$ binary Goppa code $C(L, x^9+1)$ (another best known binary Goppa code introduced in \cite{LC-84}).




\subsection{Construction of $[240,21,104]$ code:}
The binary Goppa code $C\left(L, (x^{16} + x)^6\right)$ is a $[240, 21, 104]$ code. This linear code has a higher minimum distance than the current best known $[240, 21, 98]$ binary code. Its weight enumerator polynomial is given by:
\begin{align*}
 x^{240} &+ 143640 x^{136} y^{104} + 408255 x^{128} y^{112} + 993360 x^{120} y^{120}\\
 &+ 408255 x^{112} y^{128} + 143640 x^{104} y^{136} + y^{240}.
\end{align*}

\noindent We remark that this code may also be constructed from $C\left(L,(x^{17}+1)^6\right)$ by appending an even parity check bit.




\subsection{Construction of $[241, 21, 104]$ binary code:}
The binary Goppa code $C\left(L, (x^{15} + 1)^6\right)$ is a $[241, 21, 104]$ code. This linear code has a higher minimum distance than the current best known $[241, 21,99]$ binary code. Its weight enumerator polynomial is given by:
\begin{align*}
 x^{240}y &+ 143640 x^{136} y^{105} + 408255 x^{128} y^{113} + 993360 x^{120} y^{121}\\
 &+ 408255 x^{112} y^{129} + 143640 x^{104} y^{137} + y^{241}.
\end{align*}



\subsection{Constructions of derived codes}

Recall that if $C$ is a $[n, k, d]$ code, then we can construct a $[n - 1, k, d - 1]$ by puncturing, and a $[n - 1, k - 1, d]$ code by shortening. Consider our $[239, 21, 103]$ code. By puncturing it 12 times we get best known codes with the following parameters:
\begin{align*}
    &[238, 21, 102], \, [237, 21, 101], \, [236, 21, 100], \, [235, 21, 99], \, [234, 21, 98], \, [233, 21, 97], \\
    &[232, 21, 96], [231, 21, 95], \, [230, 21, 94], \,[229, 21, 93] \, [228, 21, 92], \, \text{and} \, [227, 21, 91].
\end{align*}

\noindent By shortening our $[240, 21, 104]$ code we get a $[239, 20, 104]$ code. By puncturing this one 7 times we get codes with the following parameters:
\begin{align*}
    &[238, 20, 103], \, [237, 20, 102], \, [236, 20, 101], \, [235, 20, 100], \, [234, 20, 99], \, \\  
    &[233, 20, 98], \, \text{and} \, [232, 20, 97].
\end{align*}

\noindent By extending the $[241,21,104]$ code to further lengths we get codes with following parameters:
\[
 [242,21,104], \, [243,21,104], \, [245,21,104], \, [245,21,104], \, [246,21,104], \, \text{and} \, [247,21,104].
\]

All of these derived linear codes mentioned have higher minimum distance than the current best known codes (according to \cite{CODE-TABLE}) with the respective length and dimension. These derived codes can also be derived from the [239,21,103] binary Goppa code.

\section{Acknowledgements and other Best Known Codes}

We are very grateful to M. Grassl for pointing out the following two construtions of new best known binary codes derived from the $[240, 21, 104]$ binary Goppa code.

With the technique from \cite{GW} it is actually possible to puncture the code at suitably chosen
positions to obtain best known binary codes of parameters $[208,21,81]$, $[210,21,82]$, $[213,21,84]$, $[215,21,85]$, $[218,21,87]$, $[220,21,88]$, $[223,21,90]$, $[226,21,92]$, $[229,21,94]$ and $[229,21,94]$, but the very positions depend
on the choice of the ordering of the elements of $\F_{256}$ when constructing the Goppa code in first place.

Applying Construction X \cite{CON-X} to the $[240, 21, 104]$ binary Goppa code, we can also find a best known $[249, 21, 106]$ binary code and a $[254, 22, 106]$ best known binary code.

\end{document}